\documentclass[12pt]{article}
\usepackage{graphicx}
\begin{document}
\newcommand{\sheptitle}
{Lowering solar mixing angle in inverted hierarchy
 without   charged lepton corrections}
\newcommand{\shepauthor}
{N.Nimai Singh$^{a,}$\footnote{Regular Associate, 
The Abdus Salam ICTP, Trieste, Italy\\ E-mail: {\it nimai03@yahoo.com}},
 Monisa Rajkhowa$^{a,b}$ and Abhijit Borah$^c$}
\newcommand{\shepaddress}
{$^a$Department of Physics, Gauhati University, Guwahati-781 014, India \\
$^b$Department of Physics, Science College, Jorhat-785 010, Assam, India \\
 $^c$Department of Physics, Fazl Ali College, Mokokchung - 798 601, Nagaland, India}
\newcommand{\shepabstract}
{In the present work, the inverted hierarchical neutrino mass model which is  characterised by  opposite CP 
parity in the first two mass eigenvalues  $(m_1, -m_2, m_3)$, is studied in order  to 
lower the predicted value of solar mixing angle $\tan^2\theta_{12}$, from the  tri-bimaximal mixing (TBM), without sacrificing 
the conditions of maximal atmospheric mixing angle ($\theta_{23}=\pi/4$) 
and zero reactor angle ($\theta_{13}=0$).
 The present attempt  is different from the earlier 
 approach where the correction from the charged lepton mass matrix is included in the leptonic mixing matrix to 
lower the prediction on solar mixing angle.
The  lowering of the solar mixing angle without charged lepton correction,
 can be obtained through the  variation of the input value of a flavour twister term 
present  in the texture of  neutrino  mass matrix having a  2-3 symmetry. 
 The present analysis agrees with  the latest experimental bounds on  $\bigtriangleup m^2_{21}$ and 
$|\bigtriangleup m^2_{23}|$. It also represents an  important result on the survival of the inverted hierarchical
 neutrino mass models having  opposite CP parity in the first two eigenvalues.}

\begin{titlepage}
\begin{flushright}
hep-ph/0603154
\end{flushright}
\begin{center}
{\large{\bf\sheptitle}}
\bigskip\\
\shepauthor
\\
\mbox{}\\
{\it\shepaddress}\\
\vspace{.5in}
{\bf Abstract}
\bigskip
\end{center}
\setcounter{page}{0}
\shepabstract
\end{titlepage}
\section{Introduction}
Recent  observational data [1] on neutrino oscillations, indicates a clear departure from the tri-bimaximal mixing(TBM)
or  Harrison-Perkins-Scott(HPS) mixing pattern[2]. The most recent SNO experimental determination[3] of solar mixing angle leads to 
$\tan^2\theta_{12}=0.45^{+0.09}_{-0.08}$ compared with $\tan^2\theta_{12}=0.5$ in HPS scheme. So far there is no strong claim 
for a clear departure from the maximal atmospheric mixing angle ($\tan^2\theta_{23}=1.0$) and zero value of
 CHOOZ angle ($\sin\theta_{13}=0$),
though  only upper bound for $\sin\theta_{13}$ has been known at the moment. Future measurements
 may find  a very small value of $\sin\theta_{13}$ which can be approximated by zero[4]. Such possibility  does not yet contradict with 
the non-observation of Dirac CP phase angle. Relevance of exact zero value of CHOOZ mixing angle is gaining momentum in the literature, 
and there are several discussions[5] on the experimental setup for mass hierarchy measurements under  $\sin\theta_{13}=0$ condition.

Since the present data on solar and atmospheric neutrino oscillation experiments give only the mass-square differences,
 we usually have three models of neutrino mass levels[6], (a) degenerate model:
 $m_1\sim m_2\sim m_3 >>\sqrt{\bigtriangleup m^2_{23}}$, 
(b) inverted hierarchical model: $m_1\sim m_2>>m_3$ with $\bigtriangleup m^2_{23} = m^2_3-m^2_2<0$ 
 and $m_{1,2}\sim \sqrt{\bigtriangleup m^2_{23}}$, and (c) normal hierarchical model: $m_1<<m_2<<m_3$
 with $\bigtriangleup m^2_{23}>0$ and 
$m_3\sim \sqrt{\bigtriangleup m^2_{23}}$. Relative phase of   CP parity between  these mass eigenvalues will
 again lead to further subdivisions
in these three categories of  neutrino mass models[6]. 

Inspite of inferences drawn from the presently  available neutrino experiments[1],
 the current scenario concerning the pattern of the three neutrino masses, is not a clear one[7,8], 
though some reactor-based experiments are trying to measure  the correct pattern of neutrino masses[9].
 In this  context,  the inverted hierarchical model having 
opposite CP parity in first two mass eigenvalues ($m_1,-m_2,m_3$), has been given some special attention due to  
its high stability under radiative corrections in  the minimal supersymmetric standard model(MSSM)[10-12], and 
 considerable theoretical effort  has been made in this direction during the last one decade[13].

The  origin of the inverted hierarchical mass model  can be routed through the exact global charge ($L_e$-$L_\mu$-$L_\tau$) symmetry, 
and its subsequent breaking due to possible quantum gravity corrections or some flavour $U(1)_F$ 
symmetry violation[14]. A general form of inverted hierarchical mass matrix  having a 2-3 symmetry in 
$\nu_\mu$$\leftrightarrow$$\nu_\tau$ sector, can be written  as
\begin{equation}
m_{LL}=\left(\begin{array}{ccc}
\delta_{1}& 1& 1 \\
1& \delta_{2}& \delta_{3} \\
1& \delta_{3}& \delta_{2}
\end{array}\right)m_{0}
\end{equation}
where the perturbation parameters $\delta_{1}$, $\delta_{2}$, $\delta_{3}$ are assumed to be  smaller than unity, 
and $m_{0}$ is the  input quantity representing neutrino mass scale[11]. 
Such form of  neutrino mass matrix can  be successfully generated[15] by  using the celebrated  see-saw formula.

The conditions of maximal atmospheric mixing ($\tan^{2}\theta_{23}$ = 1)
 and  $\sin\theta_{13}$=0, are the general consequences of the  2-3 symmetry.
However, the prediction on  solar mixing angle  $\tan^{2}\theta_{12}$, is nearly maximal ($\tan^{2}\theta_{12}\sim 1$).
Several attempts[14-16]  have been  made  to tone down the maximal solar mixing angle, while  keeping $\sin\theta_{13}$ 
within the observational bound. 
In particular, the correction, $U_{eL}$  from the diagonalization of the charged lepton mass matrix 
with  non-zero texture in 1-2 block, in the leptonic mixing matrix, $U_{MNS}$=$U_{\nu L}^\dag U_{eL}$, 
can  reduce the solar mixing angle from its maximal value[11]. 
However, this correction on solar angle is strongly constrained by the observed smallness of $\sin\theta_{13}$, 
as the amount of deviation 
from a maximal solar angle is of the order of $\theta_{13}$. 
The mechanism can only work[5] if the value of $\theta_{13}$ is very close
 to its present upper bound, $\sin\theta_{13} < 0.16$.
Since the upper bound of $\sin\theta_{13}$ is known, $\sin\theta_{13}$ may be very ``small'' or
  zero exactly[4,5].

In the present work, we   propose a particular texture of the inverted hierarchical neutrino mass model, 
which 
 has the potential to lower the solar mixing angle 
from its tri-bimaximal value without affecting the maximal atmospheric mixing ($\tan^{2}\theta_{23}$=1) and 
 $\sin\theta_{13}$=0 conditions. In section 2, we present a parametrisation of neutrino mass matrix in terms of a flavour
 twister which is important for giving deviation from tribimaximal mixings. Section 3 is devoted to numerical analysis using MATHEMATICA
 and main results of the investigation. In section 4 we conclude with a summary and discussion.

\section{Parametrisation in term of flavour twister and deviation from tribimaximal mixings}
We diagonalise  the neutrino  mass matrix $m_{LL}$ in equation (1). We get the following mass eigenvalues,
\begin{eqnarray}
m_{1,2}=\frac {m_{0}}{2}[ (\delta_{1}+\delta_{2}+\delta_{3})\pm x ],\ \
m_{3}=m_{0}[ (\delta_{2}-\delta_{3})];
\end{eqnarray}
\begin{equation}
x^{2}=8+(\delta_{1}^{2}+ \delta_{2}^{2}+ \delta_{3}^{2}) - 
{2}\delta_{1}\delta_{2} -{2}\delta_{1}\delta_{3}+{2}\delta_{2}\delta_{3}.
\end{equation}
The three  mixing angles are given by 
\begin{eqnarray}
\tan^{2}\theta_{23}={1}, \ \ \sin\theta_{13}=0, \ \
\tan{2}\theta_{12}=\frac{2\sqrt{2}}{(\delta_{1}-\delta_{2}-\delta_{3})}.
\end{eqnarray}
In general, any mass matrix of the type in eqation (1) having 2-3 symmetry, can be diagonalized by 
the following  unitary matrix ( $\theta_{23}$ =$\pi/4$, $\theta_{13}$=0) [4,17],
\begin{equation}
U=\left(\begin{array}{ccc}
\cos\theta_{12}& -\sin\theta_{12}& 0 \\
\frac{\sin\theta_{12}}{\sqrt{2}}&\frac{\cos\theta_{12}}{\sqrt{2}} & -\frac{1}{\sqrt{2}} \\
\frac{\sin\theta_{12}}{\sqrt{2}}&\frac{\cos\theta_{12}}{\sqrt{2}} & \frac{1}{\sqrt{2}}
\end{array}\right)
\end{equation}
where the solar angle  $\theta_{12}$ in eq.(4)  is arbitrary, but it  can  be fixed by input values $\delta_{{1}, {2}, {3}}$ 
appearing in the texture of neutrino mass matrix (1). 
 The  tri-bimaximal mixing(TBM) or Harrison-Perkinson-Scott (HPS)
 mixing pattern[2] is thus obtained  when we choose $\cos\theta_{12}$=$\sqrt\frac{2}{3}$,
 and $\sin\theta_{12}$=$\frac{1}{\sqrt{3}}$,
 leading to  $\tan^{2}\theta_{12}$=$\frac{1}{2}$ (or $\tan 2\theta_{12}=2\sqrt 2$), and it assumes the following form,
\begin{equation}
U_{TBM}=\left(\begin{array}{ccc}
\sqrt{\frac{2}{3}}& -\frac{1}{\sqrt{3}}& 0 \\
\frac{1}{\sqrt{6}}&\frac{1}{\sqrt{3}} & -\frac{1}{\sqrt{2}} \\
\frac{1}{\sqrt{6}}&\frac{1}{\sqrt{3}} & \frac{1}{\sqrt{2}}
\end{array}\right)
\end{equation}

For diagonal  neutrino mass matrix $D=diag(m_1, m_2, m_3)$, the mass matrix $m_{LL}=U_{TMB}DU^{\dagger}_{TBM}$ 
generally takes a simple form, 
\begin{equation}
m_{LL}=\left(\begin{array}{ccc}
A& B& B \\
B& A-C& B+C \\
B& B+C& A-C
\end{array}\right)m_{0}
\end{equation}
where the elements are expressible in terms of linear combinations of three masses.
 This particular form of mass matrix (7) is a consequence of tribimaximal mixings and  can also be  derived from the general 
 $S_3$ symmetry[18]. It puts certain contraints for tribimaximal mixings on the mass matrix of  eq.(1). From  eq.(4) we have 
$(\delta_1-\delta_2-\delta_3)=\pm1$ for tri-bimaximal mixings, and this condition requires only 
two parameters in term of a flavour twister[19]. Thus the condition 
of tribimaximal mixings and deviations from it can be expressed in term of flavour twister $\eta /\epsilon$,
   
$(\delta_1-\delta_2-\delta_3)=(2-\eta /\epsilon)$.

 This condition can be easily satisfied by the choice of a simple parametrisation of 
$\delta_{1}$,  $\delta_{2}$,  $\delta_{3}$,  $m_{0}$ in terms of only 
 two new  parameters $\eta$ and $\epsilon$, 
\begin{equation}
\delta_{1}=2(1-\frac{1}{2\epsilon}), \ \ 
\delta_{2}=-\frac{1}{2\epsilon}, \ \ 
\delta_{3}=(\frac{\eta}{\epsilon}-\frac{1}{2\epsilon}), \ \ 
m_{0}=(0.05\times\epsilon)eV
\end{equation}
Substituting $(8)$ in eq.$(1)$ ,
\begin{equation}
m_{LL}=
\left( \begin{array}{ccc}
1-2\epsilon & -\epsilon & -\epsilon\\
-\epsilon & \frac{1}{2} &  \frac{1}{2}-\eta \\
-\epsilon & \frac{1}{2}-\eta  &  \frac{1}{2}\\
\end{array}\right)m_{0}^{'}
\end{equation}
where, $m_{0}^{'}=0.05eV$.
The mass eigenvalues and mixing angles of (9) are,
\begin{equation}
m_{1,2}=\frac{m_{0}^{'}}{2}[2-2\epsilon-\eta \pm x],\ \ \
m_{3}=\eta m_{0}^{'};
\end{equation}
\begin{equation}
x^2=12\epsilon^{2}-4\eta\epsilon+\eta^{2};
\end{equation}
\begin{equation}
\tan2\theta_{12}= \frac{2\sqrt2}{2-\frac{\eta}{\epsilon}};
\end{equation}
\begin{equation}
\tan^2\theta_{23}=1;\ \ \
\sin\theta_{13}=0.
\end{equation}
The solar and atmospheric mass-square differences can be expressed as,
\begin{equation}
\bigtriangleup m_{21}^2=[m_2^2-m_1^2], \ \ \  \\
\bigtriangleup m_{23}^2=[m_2^2-m_3^2].
\end{equation}
with the latest observational bounds,
\begin{equation}
\bigtriangleup m_{21}^2=(7.2-9.5)\times10^{-5}eV^2, \ \ \  \bigtriangleup m_{23}^2=(1.28-4.2)\times10^{-3}eV^2
\end{equation}
By convention $\bigtriangleup m_{21}^2$ is always defined as positive, $m_2^2\geq m_1^2$, for $\tan^2\theta_{12}<1$.
From eq.$(12)$, it is clear that $\tan^2\theta_{12}$ can be fixed by proper choice of the  flavour twister 
term $\frac{\eta}{\epsilon}$.

\section{Numerical analysis and results}
Both graphical and numerical methods are employed to find the  value of $\eta$ (or $\epsilon$) for a fixed $\tan^2\theta_{12}$.
 We first graphically solve eqs. $(12),(14),(15)$ using Mathematica Programmes(MATHEMATICA 5).
 Numerical solution of eq.$(14)$  leads to  exact  values of $\eta$ (or $\epsilon$)  consistent with 
 $\bigtriangleup m_{21}^2$  and  $\bigtriangleup m_{23}^2$ in the data (15). 
The process requires overall consistent approach to pindown the exact values of $\eta$ (or $\epsilon$) 
for the given $\tan^2\theta_{12}$ value.  

For demonstration  we consider the case for $\tan^2\theta_{12}=0.5$. Two different  values of $\frac{\eta}{\epsilon}$ 
satisfying $\tan^2\theta_{12}=0.5$ are $\frac{\eta}{\epsilon}=1$ and $\frac{\eta}{\epsilon}=3$. From the graphical 
solution of  eqs. $(12),(14),(15)$, it is found that there are two positive ranges of values 
of $\eta$ for  $\frac{\eta}{\epsilon}=1$ and one negative range  $\frac{\eta}{\epsilon}=3$ respectively,
 for the given $\tan^2\theta_{12}$.  The three allowed ranges of $\eta$ are tabulated as,\\
\begin{center}
\begin{tabular}{|r|c|l|}
\hline 
$\frac{\eta}{\epsilon}$ & $\tan^2\theta_{12}$ & range of $\eta$ \\
\hline 
1 & 0.5 & 0.6603-0.6618\\
1 & 0.5 & 0.0048-0.0064 \\
3 & 0.5 & -0.0187--0.0142\\
\hline
\end{tabular}
\end{center}
We are interested only in the first range in the present analysis, as this range leads to neutrino mass model having opposite CP parity in 
the first two mass eigenvalues. The other two ranges of $\eta$ predict even CP parity in mass eigenvalues.
For deviation from tribimaximal mixings, we also find out the ranges of  
 $\eta$ for $\tan^2\theta_{12}=0.45, 0.35$ using  the same procedure and results are  presented in Table-1.

\begin{center}
{\bf Table-1:} Lowering of $\tan^{2}\theta_{12}$$\leq 0.5$ for fixed $\tan^{2}\theta_{23}=1$, 
$\sin\theta_{13}=0$.  \\
\begin{tabular}{|r|c|c|c|l|}
\hline 
$\frac{\eta}{\epsilon}$ & $\tan^2\theta_{12}$ & range of $\eta$ & $\bigtriangleup m_{21}^2(10^{-5}eV^2)$ & $\bigtriangleup m_{23}^2(10^{-3}eV^2)$\\
\hline 
1 & 0.5 & 0.6603-0.6618 & 9.499-7.199 & 1.410-1.405 \\
0.8405 & 0.45 & 0.5865-0.5878 & 9.500-7.200 & 2.032-2.029\\
0.4462 & 0.35 & 0.3622-0.3628 & 9.430-7.200 & 4.010-4.008\\
\hline
\end{tabular}
\end{center}  

 It is seen that the quantity  $\bigtriangleup m_{21}^2$ is quite sensitive to the corresponding change in the input values
 of $\eta$, whereas such variation is almost absent in the prediction of $\bigtriangleup m_{23}^2$.
\begin{figure}
\begin{center}\includegraphics[%
            width=0.50\textwidth]{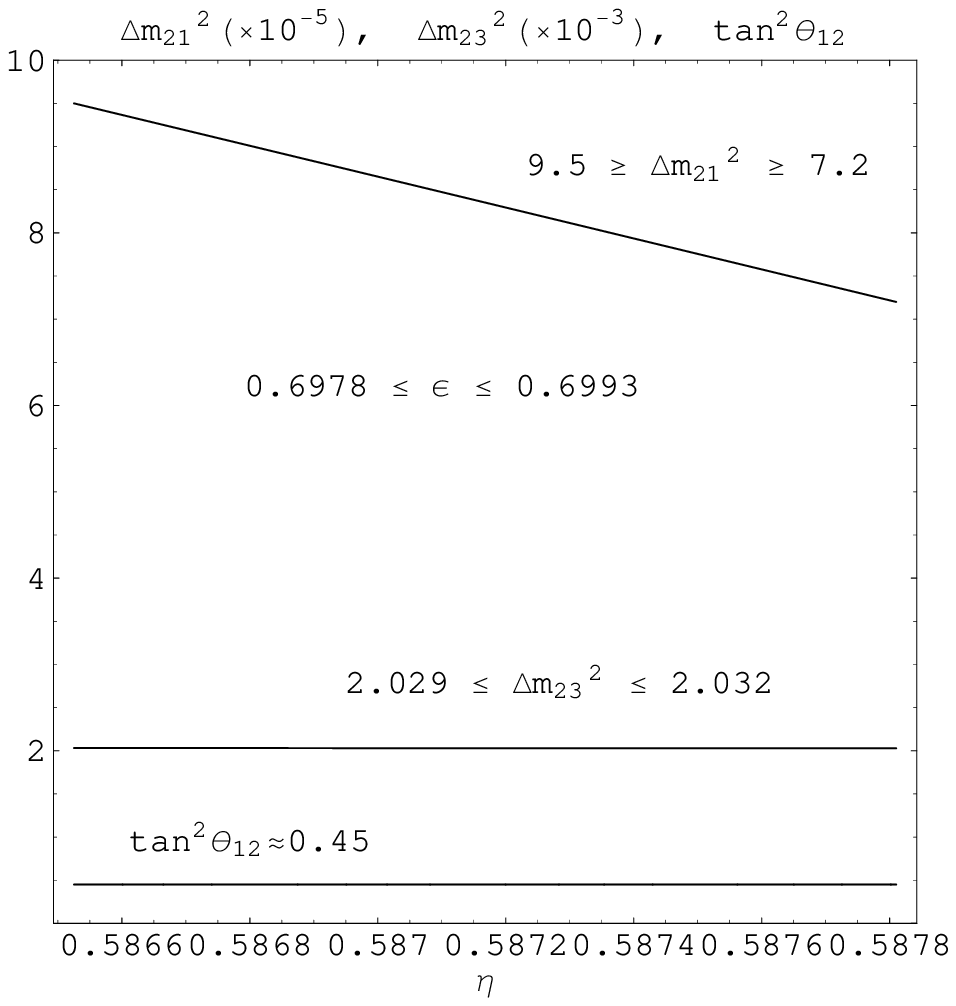}\hfill{}\includegraphics[%
            width=0.50\textwidth]{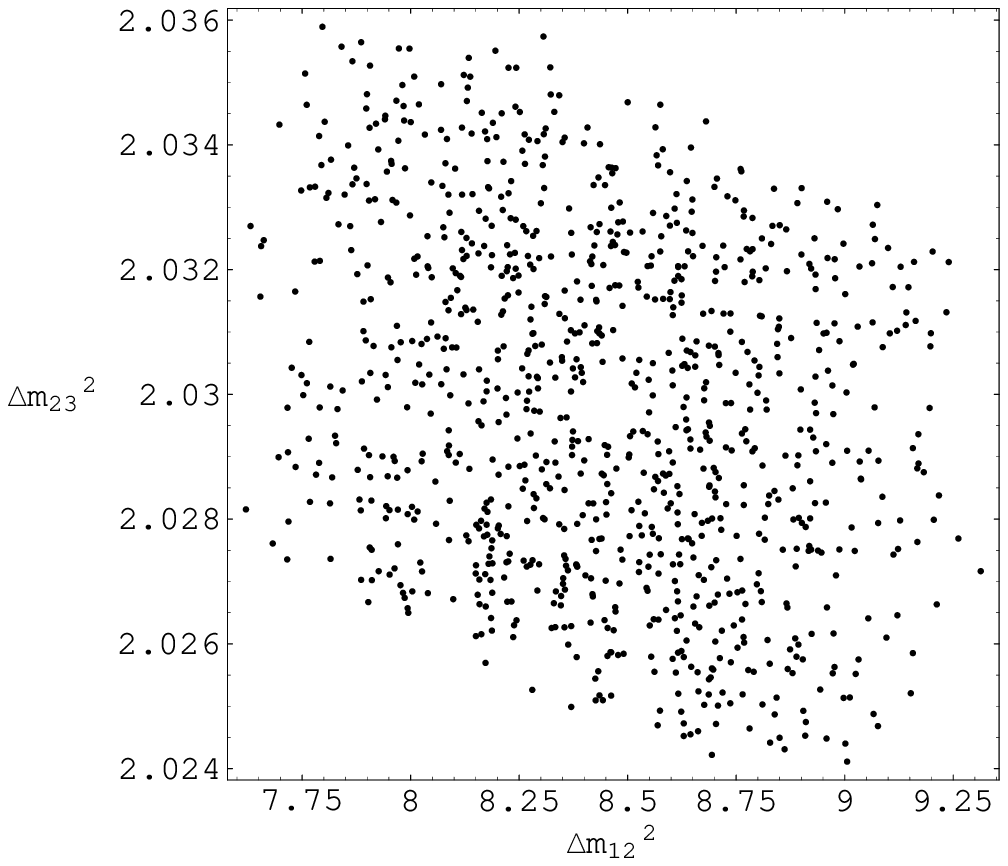}
\end{center}
\caption{Predictions on  $\Delta m^{2}_{21}$ in the unit $(10^{-5}eV^2)$ and 
 $\Delta m^{2}_{23}$ in the unit  $(10^{-3}eV^2)$ for the value  $\tan^2\theta_{12}=0.45$ in the
  range $0.5865 \leq \eta \leq 0.5878$ and the corresponding correlation graph.}
\end{figure}

We  also present the graphical solution for the case $\tan^2\theta_{12}=0.45$  along with a  correlation plot between 
 $\Delta m^{2}_{21}$ and $\Delta m^{2}_{23}$
in Fig.1. Similar diagrams and correlation plots for other two  remaining cases in Table-1, can in principle 
be drawn in the same fashion. 

For completeness,  we also give other two possible solutions of flavour twister $\eta/\epsilon$ for $\tan^2\theta_{12}=0.5, 0.45$,
which predict even CP parity in the first two mass eigenvalues in Table-2. Such models are not stable under 
radiative corrections in MSSM. However it is interesting to note that these two types of inverted hierarchical models have 
the same origin with the same  form of mass matrix.

{\bf Table-2:} Lowering of $\tan^{2}\theta_{12}$$\leq 0.50$ for fixed $\tan^{2}\theta_{23}=1$, 
$\sin\theta_{13}=0$ for different ranges  of $\eta$(or $\epsilon$) which lead to even CP parity in mass eigenvalues.
\begin{center}
\begin{tabular}{|l|l|l|l|r|}
\hline
$\frac{\eta}{\epsilon}$ &  $\tan^{2}\theta_{12}$ &
 range of $\eta$ &  $\Delta m^{2}_{21} ({10}^{-5} eV^{2})$  
&   $\Delta m^{2}_{23} ({10}^{-3} eV^{2})$\\ \hline 
1.00 & 0.50 & 0.0048 - 0.0064 & 7.200 - 9.499  & 2.499 - 2.499 \\
0.84 & 0.45 & 0.0040 - 0.0053 & 7.173 - 9.499 & 2.502 - 2.503\\
3.00 & 0.50 & -0.0187 - -0.0142 & 9.499 - 7.199 & 2.630 - 2.595\\
3.16 & 0.45 & -0.0193 - -0.0147 & 9.500 - 7.199 & 2.626 - 2.596 \\
\hline
\end{tabular}
\end{center}

\section{Summary and conclusion}
The inverted hierarchical neutrino mass  model characterised by  opposite 
CP parity in the  first two mass eigenvalues $(m_1, -m_2, m_3)$, 
is generally  stable under renormalization group evolution in MSSM, but its predicted value of  
solar mixing angle, is generally larger than the observed  data. The usual approach to lower the prediction of solar mixing 
angle is carried out through the  correction from charged lepton mass matrix. The texture of neutrino mass matrix 
studied  in the present work, has the potential to predict lower values of solar mixing angle 
beyond its  tribimaximal value without taking usual correction from charged lepton sector.  
This is carried out by  changing  the input value of a flavour twister 
 present  in the mass matrix[19].
The presence of a  2-3 symmetry in the mass matrix, protects   the conditions for 
$\tan^{2}\theta_{23}=1$  and $\sin\theta_{13}=0$.

  The  analysis presented here though phenomenological, supports  the survival of the inverted hierarchical 
 neutrino mass model with opposite CP parity in the first two mass eigenvalues. 
This model is  not yet ruled out  and this competes with the normal hierarchical model at equal footing.
Future neutrino experiments[5,20] are expected 
 to  confirm the correct pattern of neutrino mass hierarchy and  will be capable of probing very low $\theta_{13}$.
 The present work may be useful  to model building on tri-bimaximal mixings(TBM)
and possible deviations based on discrete as well as non-abelian gauge groups[21].

\section*{Acknowledgement}One of us (NNS) is thankful to Prof. Guido  Altarelli and Prof. Ernest Ma for 
useful interactions during WHEPP-9, held at  Institute of Physics, Bhubaneswar, India during Jan,3-14, 2006.
Monisa Rajkhowa acknowledges the UGC  for awarding fellowship under fellowship improvement programme (FIP).

\end{document}